
\magnification=1200
\voffset=0 true mm
\hoffset=0 true in
\hsize=6.5 true in
\vsize=8.5 true in
\normalbaselineskip=13pt
\def\doublespace{\baselineskip=20pt plus 3pt\message{double space}}
\def\singlespace{\baselineskip=13pt\message{single space}}
\let\spacing=\singlespace
\parindent=1.0 true cm



\newcount\equationumber \newcount\sectionumber
\sectionumber=1 \equationumber=1
\def\setsection{\global\advance\sectionumber by1 \equationumber=1}

\def\numbe{{{\number\sectionumber}{.}\number\equationumber}
                            \global\advance\equationumber by1}
\def\numberit{\eqno{(\number\equationumber)} \global\advance\equationumber by1}

\def\numberal{(\number\equationumber)\global\advance\equationumber by1}

\def\sectionit{\eqno{(\numbe)}}

\def\ccf#1{\,\vcenter{\normalbaselines
    \ialign{\hfil$##$\hfil&&$\>\hfil ##$\hfil\crcr
      \mathstrut\crcr\noalign{\kern-\baselineskip}
      #1\crcr\mathstrut\crcr\noalign{\kern-\baselineskip}}}\,}
\def\scf#1{\,\vcenter{\baselineskip=9pt
    \ialign{\hfil$##$\hfil&&$\>\hfil ##$\hfil\crcr
      \vphantom(\crcr\noalign{\kern-\baselineskip}
      #1\crcr\mathstrut\crcr\noalign{\kern-\baselineskip}}}\,}

\def\small3j#1#2#3#4#5#6{\def\st{\scriptstyle} 
   \bigl(\scf{\st#1&\st#2&\st#3\cr
           \st#4&\st#5&\st#6\cr} \bigr)}




\def\ref#1{$^{#1)}$}


\def\upa#1{\raise 1pt\hbox{\sevenrm #1}}
\def\dna#1{\lower 1pt\hbox{\sevenrm #1}}
\def\dnb#1{\lower 2pt\hbox{$\scriptstyle #1$}}
\def\dnc#1{\lower 3pt\hbox{$\scriptstyle #1$}}
\def\upb#1{\raise 2pt\hbox{$\scriptstyle #1$}}
\def\upc#1{\raise 3pt\hbox{$\scriptstyle #1$}}
\def\hprime{\raise 2pt\hbox{$\scriptstyle \prime$}}
\def\ccom{\,\raise2pt\hbox{,}}


\def\asymptotically#1{\;\rlap{\lower 4pt\hbox to 2.0 true cm{
    \hfil\sevenrm  #1 \hfil}}
   \hbox{$\relbar\joinrel\relbar\joinrel\relbar\joinrel
     \relbar\joinrel\relbar\joinrel\longrightarrow\;$}}
\def\Asymptotically#1{\;\rlap{\lower 4pt\hbox to 3.0 true cm{
    \hfil\sevenrm  #1 \hfil}}
   \hbox{$\relbar\joinrel\relbar\joinrel\relbar\joinrel\relbar\joinrel
     \relbar\joinrel\relbar\joinrel\relbar\joinrel\relbar\joinrel
     \relbar\joinrel\relbar\joinrel\longrightarrow$\;}}

\def\dal{\mathop{\sqcup\hskip-6.4pt\sqcap}\nolimits}

\catcode`@=11
\def\C@ncel#1#2{\ooalign{$\hfil#1\mkern2mu/\hfil$\crcr$#1#2$}}
\def\gf#1{\mathrel{\mathpalette\c@ncel#1}}      
\def\Gf#1{\mathrel{\mathpalette\C@ncel#1}}      

\def\gapx{\lower 2pt \hbox{$\buildrel>\over{\scriptstyle{\sim}}$}}
\def\lapx{\lower 2pt \hbox{$\buildrel<\over{\scriptstyle{\sim}}$}}

\def\nablaleft{\hbox{\raise 6pt\rlap{{\kern-1pt$\leftarrow$}}{$\nabla$}}}
\def\nablaright{\hbox{\raise 6pt\rlap{{\kern-1pt$\rightarrow$}}{$\nabla$}}}
\def\nablaboth{\hbox{\raise 6pt\rlap{{\kern-1pt$\leftrightarrow$}}{$\nabla$}}}

\def\boks#1#2{{\hsize=#1 true cm\parindent=0pt
  {\vbox{\hrule height1pt \hbox
    {\vrule width1pt \kern3pt\raise 3pt\vbox{\kern3pt{#2}}\kern3pt
    \vrule width1pt}\hrule height1pt}}}}

\def\heading{ }
\def\range{ }

\def\body{\vfill\eject\parindent=1.0 true cm
 \ifx\spacing\singlespace\singlespace\else\doublespace\fi}
\def\title#1{\centerline{{\bf #1}}}

\def\today{\ifcase\month\or
  January\or February\or March\or April\or May\or June\or
  July\or August\or September\or October\or November\or December\fi
  \space\number\day, \number\year}
\let\date=\today
\newcount\hour \newcount\minute
\countdef\hour=56
\countdef\minute=57
\hour=\time
  \divide\hour by 60
  \minute=\time
  \count58=\hour
  \multiply\count58 by 60
  \advance\minute by -\count58

\def\sectionskip{\penalty-500\vskip24pt plus12pt minus6pt}

\def\sec{\hbox{\lower 1pt\rlap{{\sixrm S}}{\hbox{\raise 1pt\hbox{\sixrm S}}}}}
\def\section#1\par{\goodbreak\message{#1}
    \sectionskip\nobreak\noindent{\bf #1}\vskip0.3cm \noindent}

\nopagenumbers
\headline={\ifnum\pageno=\count31\frontheadline
  \else{\ifnum\pageno=0\frontheadline
     \else{{\raise 2pt\hbox to \hsize{\paperhead}}}\fi}\fi}

\footline={\centerline{\sevenbf \folio}}
\def\frontheadline{\sevenbf \hfil}
\def\paperhead{\sevenbf \heading\ \range\hfil\folio}
\newdimen\pagewidth \newdimen\pageheight \newdimen\ruleht
\maxdepth=2.2pt
\pagewidth=\hsize \pageheight=\vsize \ruleht=.5pt

\def\onepageout#1{\shipout\vbox{ 
    \offinterlineskip 
  \makeheadline
    \vbox to \pageheight{
         #1 
 \ifnum\pageno=\count31{\vskip 21pt\line{\the\footline}}\fi
 \ifvoid\footins\else 
 \vskip\skip\footins \kern-3pt
 \hrule height\ruleht width\pagewidth \kern-\ruleht \kern3pt
 \unvbox\footins\fi
 \boxmaxdepth=\maxdepth}
 \advancepageno}}
\output{\onepageout{\pagecontents}}
\count31=-1
\topskip 0.7 true cm
\pageno=0
\doublespace
\centerline{\bf Consistency of the Nonsymmetric Gravitational Theory}
\centerline{\bf J. W. Moffat}
\centerline{\bf Department of Physics}
\centerline{\bf University of Toronto}
\centerline{\bf Toronto, Ontario M5S 1A7}
\centerline{\bf Canada} \vskip 2 true in
\centerline{\bf Revised, February, 1994.}
\vskip 3 true in
{\bf UTPT-93-11. e-mail: Moffat@medb.physics.utoronto.ca}
\par\vfil\eject
\centerline{\bf Consistency of the Nonsymmetric Gravitational Theory}
\centerline{\bf J. W. Moffat}
\centerline{\bf Department of Physics}
\centerline{\bf University of Toronto}
\centerline{\bf Toronto, Ontario M5S 1A7}
\centerline{\bf Canada}
\vskip 0.4 true in
\centerline{\bf Abstract}

The NGT field equations with sources are expanded first about a
flat Minkowski background and then about a GR background
to first-order in the antisymmetric part of the fundamental
tensor, $g_{\mu\nu}$.  From the general, static spherically symmetric
solution of the field equation in empty space, we establish that there
are two conserved charges $m$ and $\ell^2$ corresponding to the two basic
gauge invariances of NGT. There is no direct contribution to the flux
of gravitational waves from the antisymmetric, $g_{[\mu\nu]}$, sector
in the linearized, lowest order of approximation, nor in the non-linear
theory. It is demonstrated that the flux of gravitational waves is finite in
magnitude and positive definite for solutions of the field equations which
satisfy the boundary condition of asymptotic flatness.
\par\vfil\eject
\proclaim 1. {\bf Introduction} \par
\vskip 0.2 true in
An analysis of the properties of gravitational radiation in the
nonsymmetric gravitational theory (NGT) is given, based on the
field equations with sources.  In previous work$^{1}$, an analysis
of the radiation problem in NGT, showed that the flux of
gravitational radiation was the same as the leading order
contribution in general relativity (GR).  This result was
shown to follow from the axisymmetric, time-dependent vacuum field
equations in a first-order expansion about a curved GR
background$^{2}$.  From an expansion of the exact, axisymmetric
time-dependent field equations in inverse powers of r, the same
result was obtained$^{3}$.

In the following, we shall analyze the problem using the NGT field
equations with sources.  In Sect. 2, we review the Lagrangian density and the
field equations of NGT and in Sect. 3, we consider the results that follow from
an
expansion of the theory about Minkowski spacetime.  Then in Sect. 4, we
study the consequences that follow from the general, static spherically
symmetric solution of the empty space field equations. We find that only
two conserved charges occur in NGT, the mass $m$ and $\ell^2$. These
charges are conserved by virtue of the two basic gauge invariances of NGT,
namely, diffeomorphism invariance and a $U(1)$ invariance of the Lagrangian
density. The other source tensor $T_{[\mu\nu]}$ contains only geometrical mass
coupling. Indeed, when $\ell^2=0$, NGT reduces to a purely geometrical
theory of gravity involving only the gravitational charge $m$. It is found
that in the flat space linearized theory, there is no direct contribution
to gravitational radiation in the wave-zone. Only the GR quadrupole radiation
manifests itself in this approximation.

In Sect 5, it is shown that by using a spin-projection
analysis, there are no propagating ghost modes. In Sect. 6, a
calculation of the flux of gravitational radiation for plane waves is shown
to be positive definite.

In Sect. 7, we expand the antisymmetric field equations about a
classical curved GR background.  We find that to first-order in
the skew part of $g_{\mu\nu}$, the flux of energy is finite and
positive for solutions of the field equations which satisfy the
boundary condition of asymptotic flatness.  As in the linear approximation,
only the GR quadrupole radiation makes a contribution to the flux of
gravitational waves.

An important aspect of the results confirming the physical
consistency of NGT, is that at no time is there any need for a
gauge invariance in the skew $g_{\mu\nu}$ sector of the theory.
Indeed, such a general gauge invariance does not exist in the
theory; the only invariances are general coordinate invariance and
an Abelian gauge invariance, associated with the Lagrange
multiplier field $W_\mu$.  This work demonstrates that
NGT is a consistent theory of gravity, avoiding the criticisms of Damour,
Deser and McCarthy$^{5-7}$.
\vskip 0.2 true in
\setsection\proclaim 2.
{\bf NGT Field Equations with Sources}
\par \vskip 0.2 true in
The Lagrangian density with sources, in NGT, is given by$^{8-11}$:
$$
{\cal L} = {\bf g}^{\mu\nu} R_{\mu\nu} (W) + {\cal L}_M,
\sectionit
$$
where ${\bf g}^{\mu\nu}=\sqrt{-g}g^{\mu\nu}$ and
$R_{\mu\nu}(W)$ is the NGT contracted curvature tensor:
$$
R_{\mu\nu}(W)=W^\beta_{\mu\nu,\beta} - {1\over
2}(W^\beta_{\mu\beta,\nu}+W^\beta_{\nu\beta,\mu}) -
W^\beta_{\alpha\nu}W^\alpha_{\mu\beta} +
W^\beta_{\alpha\beta}W^\alpha_{\mu\nu}, \sectionit $$ defined in
terms of the unconstrained nonsymmetric connection:
$$
W^\lambda_{\mu\nu}=\Gamma^\lambda_{\mu\nu}-{2\over
3}\delta^\lambda_\mu W_\nu,
\sectionit
$$
where $W_\mu\equiv
W^\lambda_{[\mu\lambda]} = {1\over 2}
(W^\lambda_{\mu\lambda}-W^\lambda_{\lambda\mu})$.  This equation
leads to:
$$
\Gamma_\mu=\Gamma^\lambda_{[\mu\lambda]}=0.
\sectionit
$$

The contravariant tensor $g^{\mu\nu}$ is defined in terms of the
equation:
$$
g^{\mu\nu}g_{\sigma\nu}=g^{\nu\mu}g_{\nu\sigma}=\delta^\mu_\sigma.
\sectionit
$$
The NGT contracted curvature tensor can be written
as
$$
R_{\mu\nu}(W) = R_{\mu\nu}(\Gamma) + {2\over 3}
W_{[\mu,\nu]},
\sectionit
$$
where $R_{\mu\nu}(\Gamma)$ is defined by
$$
R_{\mu\nu}(\Gamma ) = \Gamma^\beta_{\mu\nu,\beta} -{1\over
2} \left(\Gamma^\beta_{(\mu\beta),\nu} +
\Gamma^\beta_{(\nu\beta),\mu}\right) - \Gamma^\beta_{\alpha\nu}
\Gamma^\alpha_{\mu\beta} +
\Gamma^\beta_{(\alpha\beta)}\Gamma^\alpha_{\mu\nu}.
\sectionit
$$

The Lagrangian density for the matter sources is given by (G=c=1):
$$
{\cal L}_M= -8\pi g^{\mu\nu} {\bf T}_{\mu\nu} + {8\pi \over 3}
W_\mu {\bf S}^\mu.
\sectionit
$$

Our field equations are given by
$$
G_{\mu\nu} (W) = 8\pi T_{\mu\nu},
\sectionit
$$
$$
{{\bf g}^{[\mu\nu]}}_{,\nu} = 4\pi {\bf S}^\mu,
\sectionit
$$
where
$$
G_{\mu\nu} = R_{\mu\nu} - {1\over 2} g_{\mu\nu} R.
\sectionit
$$
The variation of the $W$ connection gives
$$
{{\bf g}^{\mu\nu}}_{, \sigma} + {\bf
g}^{\rho\nu} W^\mu_{\rho\sigma} + {\bf
g}^{\mu\rho}W_{\sigma\rho}^\nu - {\bf
g}^{\mu\nu}W^\rho_{\sigma\rho} + {2\over 3}\delta^\nu_\sigma {\bf
g}^{\mu\rho} W_\rho + {4\pi\over 3} ({\bf S}^\nu \delta^\mu_\sigma
- {\bf S}^\mu\delta^\nu_\sigma) = 0.
\sectionit
$$
These equations
can be written in the form:
$$
g_{\mu\nu,\sigma}-g_{\rho\nu}\Lambda^\rho_{\mu\sigma}-
g_{\mu\rho}\Lambda^\rho_{\sigma\nu}=0, \sectionit $$ where $$
\Lambda^\rho_{\mu\nu}=\Gamma^\rho_{\mu\nu}+D^\rho_{\mu\nu}.
\sectionit
$$
Here, $D^\rho_{\mu\nu}$ depends only on $S^\mu$ and
$g_{\mu\nu}$ and is defined by $$ g_{\rho\nu}D^\rho_{\mu\sigma} +
g_{\mu\rho}D^\rho_{\sigma\nu} =-{4\pi\over
3}S^\rho(g_{\mu\sigma}g_{\rho\nu}-g_{\mu\rho}g_{\sigma\nu}
+g_{\mu\nu}g_{[\sigma\rho]}).
\sectionit
$$

The variational principle yields for invariance under coordinate
transformations the four Bianchi identities:
$$
\left[{\bf g}^{\alpha\nu} G_{\rho\nu}(\Gamma) + {\bf
g}^{\nu\alpha}
G_{\nu\rho} (\Gamma)\right]_{,\alpha} + {g^{\mu\nu}}_{, \rho} {\bf
G}_{\mu\nu}(\Gamma) = 0.
\enskip
\sectionit
$$
The matter response equations are
$$ {1\over 2}\left(g_{\sigma\rho}{\bf T}^{\sigma\alpha} +
g_{\rho\sigma}{\bf T}^{\alpha\sigma}\right)_{,\alpha} - {1\over 2}
g_{\alpha\beta,\rho}{\bf T}^{\alpha\beta} + {1\over 3}
W_{[\rho,\nu]}{\bf S}^\nu = 0.
\sectionit
$$
 From the invariance
of ${\cal L}$ under the $U(1)$ gauge transformation:
$$
W_\mu^\prime = W_\mu + \lambda_{, \mu},
\sectionit
$$ we obtain from Neother's theorem the identity:
$$ {{\bf g}^{[\mu\nu]}}_{,\mu,\nu} \equiv 4\pi{{\bf S}^\mu}_{,\mu}
= 0.
\sectionit
$$

An important consideration, in determining the fundamental
properties of NGT, is the possible invariances allowed in the
theory.  The rigorous NGT Lagrangian density contains the
diffeomorphism invariance, associated with the invariance under
general coordinate transformations, which leads to the existence
of the four Bianchi identities (2.16).  The only other invariance
of the Lagrangian density is the Abelian invariance under the
gauge transformation (2.18), from which follows the one identity
(2.19).  There is no rigorous, general invariance associated with
the $g_{[\mu\nu]}$, and there are no further identities associated
with such an invariance.  However, as we shall demonstrate, there
is no need of any gauge
invariance associated with $g_{[\mu\nu]}$.  This will prove to be
true in the linear approximation to the theory, as well as in the
higher-order nonlinear theory.
\vskip 0.2 true in
\setsection\proclaim 3. {\bf Field Equations in the Linear Approximation}
\par \vskip 0.2 true in
Let us consider weak fields
in NGT, whereby we expand $g_{\mu\nu}$ about flat Minkowski
space$^{12}$:
$$
g_{\mu\nu} = \eta_{\mu\nu} + h_{\mu\nu},
\sectionit
$$
where $\vert h_{\mu\nu}\vert \ll 1$ and
$\eta_{\mu\nu}$ is the Minkowski metric:  $\eta_{\mu\nu}=(-1, -1,
-1, +1).$ We shall solve the field equations to lowest order in
$h_{\mu\nu}$.  Raising and lowering is done using $\eta_{\mu\nu}$,
so that from (2.5), we obtain
$$
g^{\mu\nu}=\eta^{\mu\nu}-h^{\mu\nu}+O(h^2),
\sectionit
$$
$$
h^{\mu\nu}=\eta^{\mu\lambda}\eta^{\sigma\nu}h_{\sigma\lambda}.
\sectionit
$$
Solving for $\Lambda^\lambda_{\mu\nu}$ and
$D^\lambda_{\mu\nu}$ using (2.13) and (2.15), we get
$$
\Gamma^\lambda_{\mu\nu}={1\over
2}\eta^{\lambda\sigma}(h_{\sigma\nu,\mu}
+h_{\mu\sigma,\nu}-h_{\nu\mu,\sigma})-{1\over
3}(\delta^\lambda_\nu {h_{[\mu\beta]}}^{,\beta}-\delta^\lambda_\mu
{h_{[\nu\beta]}}^{,\beta}).
\sectionit
$$

The full Lagrangian of the theory in the linear approximation is
given by
$$
{\cal L}={\cal L}_{GR}+{\cal L}_S,
\sectionit
$$
where
$$
{\cal L}_{GR}=-{1\over 4}h^{(\mu\nu)}\dal h_{(\mu\nu)}-{1\over
2} {h^{(\mu\sigma)}}_{,\sigma}{h_{(\mu\nu)}}^{,\nu} +{1\over
2}({h^{(\mu\alpha)}}_{,\alpha}-{1\over 2}h^{,\mu})h_{,\mu} +8\pi
h_{(\mu\nu)}T^{(\mu\nu)},
\sectionit
$$ and
$$
{\cal L}_S={1\over 2}({1\over
2}h^{[\mu\nu],\lambda}h_{[\mu\nu],\lambda}
+{h^{[\mu\sigma]}}_{,\sigma}{h_{[\mu\nu]}}^{,\nu})-{2\over 3}
{h^{[\mu\nu]}}_{,\nu} W_\mu-{8\pi\over
3}h^{[\mu\nu]}S_{[\mu,\nu]}-{8\pi\over 3}W_\mu S^\mu
$$
$$
+{4\pi\over 3}S^\mu S_\mu-S^{[\mu,\nu]}S_{[\mu,\nu]} +8\pi
h_{[\mu\nu]}T^{[\mu\nu]}.
\sectionit
$$ Here, $h=\eta^{\mu\nu}h_{\mu\nu}$ and
$\dal=\partial^\mu\partial_\mu$.
The field equations to lowest order are:
$$
{h_{[\mu\beta]}}^{,\beta}=4\pi S_\mu,
\sectionit
$$
$$ -{1\over 2}(\dal h_{\nu\mu}-{h_{(\nu\sigma),\mu}}^{,\sigma}
+h_{,\mu\nu}-{1\over 3}{h_{[\mu\sigma],\nu}}^{,\sigma} +{1\over
3}{h_{[\nu\sigma],\mu}}^{,\sigma}) =-{2\over
3}W_{[\mu,\nu]}+8\pi(T_{\mu\nu}-{1\over 2}\eta_{\mu\nu}T),
\sectionit
$$
where $T=\eta^{\mu\nu}T_{\mu\nu}$.

By choosing the coordinate conditions:
$$
{h_{(\sigma\mu)}}^{,\mu}={1\over 2}h_{,\sigma},
\sectionit
$$
the field equations become (3.8) and
$$
\dal h_{(\mu\nu)}=-16\pi {\tilde T}_{(\mu\nu)},
\sectionit
$$
$$
\dal h_{[\mu\nu]}=-{4\over
3}W_{[\mu,\nu]}-{8\pi\over 3}S_{[\mu,\nu]} +16\pi T_{[\mu\nu]},
\sectionit
$$
where
$$ {\tilde T}_{(\mu\nu)}=T_{(\mu\nu)}-{1\over 2}\eta_{\mu\nu}T.
\sectionit
$$

In the linear approximation, the skew Lagrangian ${\cal L}_S$ is
not invariant under the gauge transformation:
$$
\delta h_{[\mu\nu]}=\epsilon_{\mu,\nu}-\epsilon_{\nu,\mu}.
\sectionit
$$
However, the presence of the Lagrange multiplier $W_\mu$, in
${\cal L}_S$, is of crucial importance, for it guarantees that
unphysical ghost pole contributions will not occur in the theory.
If the Lagrange multiplier $W_\mu$ were absent, then it would be
vital to have invariance under the transformation (3.14) to avoid
any unphysical propagating modes.

It is important to distinguish between those degrees of freedom
that propagate as physical dynamical modes, and those that do not
couple dynamically to the matter sources.  We shall therefore
define longitudinal and transverse projection operators$^{1}$:
$$
P^L_{\mu\nu}={\partial_\mu\partial_\nu\over \dal},\quad
P^T_{\mu\nu}=\eta_{\mu\nu}-{\partial_\mu\partial_\nu\over \dal},
\sectionit
$$
and make the general decomposition:
$$
h_{[\mu\nu]}=\alpha_{\mu,\nu}-\alpha_{\nu,\mu}+\epsilon_{\mu\nu\kappa\lambda}
\beta^{[\kappa,\lambda]},
\sectionit
$$
and
$$
T_{[\mu\nu]}=K_{[\mu,\nu]}
+\epsilon_{\mu\nu\kappa\lambda}J^{[\kappa,\lambda]},
\sectionit
$$
where $K_\mu$ and $J^\mu$ are a vector and a pseudovector,
respectively.

We choose the gauge condition:
$$
{\alpha_\mu}^{,\mu}=0,
\sectionit
$$
and find that
$$
h_{[\mu\nu]}^{TT}=P^{T\alpha}_\mu
P^{T\beta}_\nu h_{[\alpha\beta]}
=\epsilon_{\mu\nu\kappa\lambda}\beta^{[\kappa,\lambda]},
\sectionit
$$
$$
h_{[\mu\nu]}^{LL}=P^{L\alpha}_\mu P^{L\beta}_\nu
h_{[\alpha\beta]}=0,
\sectionit
$$
$$
h_{[\mu\nu]}^{LT}=P^{L\alpha}_{[\mu}P^{T\beta}_{\nu]}h_{[\alpha\beta]}
=\alpha_{\mu,\nu}-\alpha_{\nu,\mu}.
\sectionit
$$
Using the gauge
condition ${W_\alpha}^{,\alpha}=0$, we have
$$
W^{TT}_{[\mu,\nu]}=W^{LL}_{[\mu,\nu]}=0,\quad
W^{LT}_{[\mu,\nu]}=W_{[\mu,\nu]}.
\sectionit
$$
Moreover, we find that
$$
T^{TT}_{[\mu\nu]}=\epsilon_{\mu\nu\kappa\lambda}J^{[\kappa,\lambda]},
\sectionit
$$
$$
T^{LL}_{[\mu\nu]}=0,
\sectionit
$$
$$
T^{LT}_{[\mu\nu]}=K_{[\mu,\nu]}.
\sectionit
$$
It follows that
$$
{T^{TT}_{[\mu\nu]}}^{,\nu}=0.
\sectionit
$$

The field equations with sources now become:
$$
{h^{LT}_{[\mu\beta]}}^{,\beta}=4\pi S_\mu,
\sectionit
$$
$$
\dal h_{[\mu\nu]}^{TT}=16\pi
\epsilon_{\mu\nu\kappa\lambda}J^{[\kappa,\lambda]},
\sectionit
$$
$$
\dal h_{[\mu\nu]}^{LT}=-{4\over 3}W_{[\mu,\nu]}-{8\pi\over
3}S_{[\mu,\nu]} +16\pi K_{[\mu,\nu]}.
\sectionit
$$
By using (3.18), (3.21) and (3.27), we have
$$
\dal \alpha_\mu=4\pi S_\mu.
\sectionit
$$
Then, (3.21) and (3.30) give
$$
\dal h^{LT}_{[\mu\nu]}=8\pi S_{[\mu,\nu]},
\sectionit
$$
and from (3.29), it follows that
$$
W_{[\mu,\nu]}=-8\pi S_{[\mu,\nu]}+12\pi K_{[\mu,\nu]}.
\sectionit
$$
Thus, from (3.28) and (3.31), we see that the auxiliary (Lagrange multiplier)
contribution $W_{[\mu,\nu]}$ can be eliminated from the field equations in the
presence of sources.
\vskip 0.2 true in
\setsection\proclaim 4. {\bf General Static Spherically Symmetric Solution}
\par
\vskip 0.2 true in
In order to establish unequivocally the number of conserved ``charges" in
NGT, we shall study the consequences of the general, static spherically
symmetric solution of the NGT field equations (2.9) and (2.10) in the
absence of sources: $S^\mu=T_{\mu\nu}=0$.

In the case of a static spherically symmetric field$^{13}$,
Papapetrou has derived the canonical form of $g_{\mu\nu}$:
$$
g_{\mu\nu}=\left(\matrix{-\alpha(r)&0&0&w(r)\cr
0&-\beta(r)&f(r)\hbox{sin}\theta&0\cr 0&-f(r)\hbox{sin}\theta&
-\beta(r)\hbox{sin}^2
\theta&0\cr-w(r)&0&0&\gamma(r)\cr}\right).
\sectionit
$$
A general solution of the NGT field equations in vacuum was obtained by
Vanstone$^{13}$, and is given by
$$
f+i\beta={\lambda b_1(i-b)\over 4(1+b^2)y}\hbox{sinh}^{-2}
\biggl[{\sqrt{b_1}\over 2}
(\hbox{ln y}-a)\biggr],
$$
$$
\alpha={(f^2+\beta^2)(y^\prime)^2\over \lambda y},\quad
\gamma={\ell^4+f^2+\beta^2\over f^2+\beta^2}y,\quad w
={\ell^2y^\prime\over \sqrt{\lambda}},
\sectionit
$$
where $y$ is an arbitrary function of $r$, $a$ is a complex constant,
$b_1=1+is$, and $b, s, \lambda$ and $\ell^2$ are real constants.

Consider now the asymptotically flat boundary conditions as $r\rightarrow
\infty$:
$$
\alpha\rightarrow 1,\quad \beta\rightarrow r^2,\quad \gamma\rightarrow 1,
\quad w\rightarrow 0.
\sectionit
$$
Vanstone proved that the only possibility to satisfy these conditions is for
$f$ {\it to vanish identically}. If we adopt the weaker condition that
$f/r^2\rightarrow 0$ as $r\rightarrow \infty$, and that the solution (4.2)
must reduce to the Schwarzschild solution when $g_{[\mu\nu]}=0$, then we get
$$
y=1-{2m\over r},\quad \lambda=4m^2,\quad b=0,\quad a=0.
\sectionit
$$
where the radial variable is only equivalent to the $r$ variable in the GR
Schwarzschild solution in the limit $\beta\rightarrow r^2$ as $r\rightarrow
\infty$.

We now put $\sqrt{b_1}=\mu+i\nu$, and obtain $\mu=\pm[1/2+1/2
\sqrt{(1+s^2)}]^{1/2}, \nu=\pm s[2+2\sqrt{(1+s^2)}]^{-1/2}$. Then we find that
$$
f={2m^2[\hbox{sinh}\xi\hbox{sin}\eta-s(\hbox{cosh}\xi\hbox{cos}\eta-1)]\over
(1-{2m\over r})(\hbox{cosh}\xi-\hbox{cos}\eta)^2},
\sectionit
$$
$$
\beta={2m^2[s\hbox{sinh}\xi\hbox{sin}\eta
+\hbox{cosh}\xi\hbox{cos}\eta-1]\over (1-{2m\over r})(\hbox{cosh}\xi
-\hbox{cos}\eta)^2},
\sectionit
$$
$$
\alpha=\biggl({f^2+\beta^2\over r^4}\biggr)\biggl(1-{2m\over r}\biggr)^{-1},
\sectionit
$$
$$
\gamma=\biggl(1+{\ell^2\over f^2+\beta^2}\biggr)\biggl(1-{2m\over r}\biggr),
\sectionit
$$
$$
w=\pm{\ell^2\over r^2},
\sectionit
$$
where $\xi=\mu\hbox{ln}y, \eta=\nu\hbox{ln}y$.
In the limit that $s=\ell^2=0$, we obtain the Schwarzschild solution:
$$
\alpha=\biggl(1-{2m\over r}\biggr)^{-1},\quad \beta=r^2,\quad
\gamma=1-{2m\over r}.
\sectionit
$$

Expanding (4.5)-(4.8) for small values of $s$, we get
$$
f={m^2s\over 3}\biggl(1-{2m\over r}+...\biggr),
\sectionit
$$
$$
\beta=r^2\biggl(1-{2s^2m^4\over 15r^4}+...\biggr),
\sectionit
$$
$$
\gamma=\biggl(1+{\ell^4\over r^4}+...\biggr)\biggl(1-{2m\over r}\biggr),
\sectionit
$$
$$\alpha=\biggl(1-{s^2m^4\over 45r^4}+...\biggr)\biggl(1-{2m\over r}\biggr).
\sectionit
$$

The first important consequence of the general, static spherically symmetric
solution is that {\it there are only two conserved charges} in NGT, the mass
$m$ and $\ell^2$. These are both conserved by virtue of the Bianchi
identities (2.16) or (2.17), as in GR, and the NGT conservation law (2.19)
with the identification:
$$
\ell^2=\int d^3x {\bf S}^0.
\sectionit
$$
Thus, only two gauge invariances play a role in NGT, the diffeomorphism
invariance and the invariance associated with the
NGT Lagrangian density under the $U(1)$ transformation (2.18). There is no
need for any extra gauge invariance associated with the $g_{[\mu\nu]}$ sector,
for there does not exist a conservation law of a new charge in the theory
besides $m$ and $\ell^2$.

The second significant consequence of the static solution is that in the
linear approximation, neglecting contributions of order $h_{(\mu\nu)}^2$,
we see from (4.11) that $h_{[23]}$ is zero to lowest order. From (3.28), it
follows that to lowest order of approximation (neglecting terms quadratic
and higher order in $m$) in the static limit: $h^{TT}_{[23]}=0$ and therefore
in general: $h^{TT}_{[mu\nu]}=0$. Then, to lowest order we have
$$
J^\mu=0,\quad T_{[\mu\nu]}=K_{[\mu,\nu]}.
\sectionit
$$
It now follows from (3.32) that to lowest order of approximation:
$$
W_{[\mu,\nu]}=0,
\sectionit
$$
in the wave-zone. Because $T_{[\mu\nu]}$ is a compact, localized source
tensor, then from (3.17) and (4.16), it follows that $K_{[\mu,\nu]}$ has
compact support and vanishes in the wave-zone together with $S^\mu$ and
(4.16) follows immediately.

In the wave-zone at infinity, the skew vacuum field equations are given by:
$$
{h^{LT}_{[\mu\beta]}}^{,\beta}=0,
\sectionit
$$
$$
\dal h_{[\mu\nu]}^{LT}=0.
\sectionit
$$

The solution for $h_{(\mu\nu)}$ coincides to lowest order with the
GR solution$^{14}$:
$$
h_{(\mu\nu)}({\bf x},t)=-4\int d^3x^\prime
{{\tilde T}_{(\mu\nu)}({\bf x}^\prime,t-\vert {\bf x} - {\bf
x}^\prime\vert) \over \vert{\bf x} - {\bf x}^\prime\vert}.
\sectionit
$$
To this order $h^{TT}_{[\mu\nu]}=0$ and the solution of equation (3.31) is
given by
$$
h^{LT}_{[\mu\nu]}({\bf x},t)=2\int d^3x^\prime{S_{[\mu,\nu]}
({\bf x^\prime},t-\vert{\bf x}-{\bf x}^\prime\vert)\over \vert
{\bf x}-{\bf x}^\prime\vert}.
\sectionit
$$
\vskip 0.2 true in
\setsection\proclaim 5. {\bf Spin Projection Analysis of the Linearized Theory}
\par
\vskip 0.2 true in
Let us write the skew part of the Lagrangian ${\cal L}_S$ in the
form$^{15}$:
$$
{\cal L}_S={1\over 2}\sum_{A,B}\phi_A O_{AB}\phi_B,
\sectionit
$$
where $\phi_A=(h_{[\mu\nu]},W_\mu)$
and $O_{AB}$ is the wave operator.  Using the methods of ref.
(16), we can decompose the fields into subspaces with spin-parity
$J^P$ and invert the operator $O_{AB}$ to obtain the saturated
propagator:
$$
\Pi=-\sum_{\psi_A,\phi_B}S_A O_{AB}^{-1}S_B,
\sectionit
$$
where $S_A=(L_{[\mu\nu]},S_\mu)$ and
$$
L_{[\mu\nu]}=T_{[\mu\nu]}+S_{[\mu,\nu]}.
\sectionit
$$
Expanding the operator $O_{AB}$ yields
$$
{\cal L}_S=\sum_{\psi_A,\phi_B,i,j,J^P}a_{ij}^{\psi\phi}(J^P)\psi_A
P^{\psi\phi}_{ij} (J^P)\phi_B,
\sectionit
$$
and
$$
\Pi=-\sum_{\psi_A,\phi_B,i,j,J^P}a^{-1\,\psi\phi}_{ij}(J^P)S_A
P^{\psi\phi}_{ij}(J^P)_{AB}S_B,
\sectionit
$$
where $a_{ij}^{\psi\phi}(J^P)$ are matrix coefficients.  A calculation of the
spin-projection operators yields
$$
P_{ij}(1^+)={1\over 2}(\theta_{\mu\alpha}\theta_{\nu\beta}
-\theta_{\nu\alpha}\theta_{\mu\beta}),
\sectionit
$$
$$
P_{ij}(1^-)=\left(\matrix{{1\over
2}(\theta_{\mu\alpha}\omega_{\nu\ beta}-
\theta_{\mu\beta}\omega_{\nu\alpha}-\theta_{\nu\alpha}
\omega_{\mu\beta}
+\theta_{\nu\beta}\omega_{\mu\alpha})& {1\over \sqrt{2}}({\hat
k}_\nu\theta_{\mu\alpha}-{\hat k}_\mu \theta_{\nu\alpha})\cr
{1\over \sqrt{2}}({\hat k}_\nu\theta_{\mu\alpha} -{\hat
k}_\mu\theta_{\nu\alpha})&\theta_{\alpha\beta}\cr}\right),
\sectionit
$$
$$
P_{ij}(0^+)=\omega_{\alpha\beta},
\sectionit
$$
where
$$
\theta_{\mu\alpha}=\eta_{\mu\alpha}-k_\mu k_\alpha
k^{-2},\quad \omega_{\mu\alpha}=k_\mu k_\alpha k^{-2},\quad {\hat
k}_\mu=k_\mu(k^2)^{-1/2}.
\sectionit
$$

These operators are orthonormal and complete within
$(h_{[\mu\nu]},W_\mu)$.  The Lagrangian ${\cal L}_S$ now yields
the matrix coefficients:
$$
a(1^+)={1\over 2}k^2,
\sectionit
$$
$$
a_{ij}(1^-)=\left(\matrix{k^2&-{\sqrt{2}i\over 3}(k^2)^{1/2}\cr
{\sqrt{2}i\over 3}(k^2)^{1/2}&0\cr}\right),
\sectionit
$$
$$
a(0^+)=0.
\sectionit
$$
The result (4.12) follows from the gauge
invariance under the transformation (2.18) and the conservation
equation (2.19).  Inverting (4.10) and (4.11) yields
$$
a^{-1}(1^+)={2\over k^2},
\sectionit
$$
$$
a^{-1}_{ij}(1^-)=\left(\matrix{0&-{3i\over \sqrt{2}}(k^2)^{1/2}\cr
{3i\over \sqrt{2}}(k^2)^{1/2}& -{9\over 2}k^2\cr}\right){1\over
k^2}.
\sectionit
$$
A calculation determines the saturated
propagator to be of the form:
$$
\Pi=-{2\over
k^2}T^{[\mu\nu]}P(1^+)_{\mu\nu\alpha\beta}T^{[\alpha\beta]}
+{3i\over
2k^2}S^\alpha(k_\nu\theta_{\mu\alpha}-k_\mu\theta_{\nu\alpha})
L^{[\mu\nu]}
$$
$$
-{3i\over 2k^2}L^{[\mu\nu]}
(k_\mu\theta_{\nu\alpha}-k_\nu\theta_{\mu\alpha})S^\alpha +{9\over
2}(S^\alpha\theta_{\alpha\beta}S^\beta).
\sectionit
$$
Here, we
have used the condition $k^\alpha\theta_{\alpha\beta}=0$ to remove
terms of the form $P(1^+)_{\mu\nu\alpha\beta}\break
S^{[\alpha,\beta]}$.

We see that the massless spin $1^-$ ghost particles do not
propagate, and only contact terms occur in this sector, whereas
the spin $1^+$ sector is neither a ghost particle nor a tachyon.
For the {\it real} version of NGT, we see that the theory is
completely ghost free, even with a nonvanishing $T_{[\mu\nu]}$.

The nonpropagating spin $1^-$ sector corresponds to the
$h^{LT}_{[\mu\nu]}$ components, while the propagating spin $1^+$
sector corresponds to the transverse $h^{TT}_{[\mu\nu]}$
components of $h_{[\mu\nu]}$.  In Sect.  6, we will prove that the
$h^{LT}_{[\mu\nu]}$ do not propagate in the wave-zone.  Although
the skew field equations are not invariant under the
transformation (3.14), there does exist a restricted gauge
invariance under the transformations:
$$
\delta
h^{LT}_{[\mu\nu]}=\epsilon_{\mu,\nu}-\epsilon_{\nu,\mu},\quad\delta
h^{TT}_{[\mu\nu]}=\epsilon_{\mu\nu\kappa\lambda}\zeta^{[\kappa,\lambda]},
\sectionit
$$
provided that $\epsilon_\mu$ and $\zeta_\mu$
satisfy:
$$
\dal \epsilon_\mu=0,\quad
{\epsilon_\mu}^{,\mu}=0,\quad \dal \zeta_\mu=0,\quad
{\zeta_\mu}^{,\mu}=0.
\sectionit
$$
This restricted gauge
invariance reduces the three degrees of freedom of the propagating
spin $1^+$ sector to one spin $0^+$ degree of freedom.  It should
be stressed at this point that we do not need to use this
restricted gauge invariance to remove potential ghost modes in the
linear approximation, for such ghost modes simply do not exist.

A similar result occurs for a spin $1^-$ field $A_\mu$ with a
scalar Lagrange multiplier $\phi$, if we replace $h_{[\mu\nu]}$
with $A_\mu$ and $W_\mu$ with $\phi\,\,$$^{18}$.  The spin $0^+$
ghost particle becomes nonpropagating because of the Lagrange
multiplier $\phi$.

If we had included a mass term in ${\cal L}_S$ of the form $d{\bf
g^{[\mu\nu]}} g_{[\mu\nu]}$, then we would be forced to have an
unphysical particle spectrum, which contains either ghosts $(d<0)$
or tachyons $(d>0)$.  Only for the massless theory are we
guaranteed the absence of ghosts, due to the presence of the
Lagrange multiplier $W_\mu$. The suggestion by Damour, Deser and McCarthy
$^{5-7}$ that a mass term be introduced into NGT to avoid bad
asymptotic behavior would necessarily lead to an unphysical
theory.
\par\vfil\eject
\setsection\proclaim 6. {\bf Plane Wave Solutions in the Linear Approximation}
\par
\vskip 0.2 true in
The plane wave solutions for $h_{(\mu\nu)}$ are
identical to those in GR.  The solution of Eqs. (4.19) is
$$
h^{LT}_{[\mu\nu]}=e^{LT}_{[\mu\nu]}\hbox{exp}(ik_\lambda
x^\lambda) +e^{*LT}_{[\mu\nu]}\hbox{exp}(-ik_\lambda x^\lambda),
\sectionit
$$
where $k^\mu k_\mu=0$,
$$
k^\mu e_{[\mu\nu]}=0,
\sectionit
$$
and to this order $h^{TT}_{[\mu\nu]}=0$.

Let us now calculate the energy-momentuum tensor of the plane
waves.  This is done using the energy-momentum pseudotensor
$t_{\mu\nu}$ of NGT.  We have to lowest order$^{12}$:
$$
t_{\mu\nu}={1\over
8\pi}(R^{(2)}_{\mu\nu}(\Gamma)-{1\over 2}\eta_{\mu\nu}
R^{(2)}(\Gamma)), \sectionit
$$
where $R^{(2)}_{\mu\nu}(\Gamma)$
is $R_{\mu\nu}(\Gamma)$ to $O(h^2)$.  We only need the average
expression, $<t_{(\mu\nu)}>$, since this is what is measured.
Averaging over space and time in a region much larger than ${\vert
{\bf k}\vert}^{-1}$, we have $$ <t_{(\mu\nu)}>={k_\mu k_\nu\over
16\pi}(e^{(\beta\gamma)}e^*_{(\beta\gamma)} -{1\over 2}\vert
\eta^{\beta\gamma}e_{(\beta\gamma)}\vert^2
+e^{[\beta\gamma]}e^*_{[\beta\gamma]}).  \sectionit $$ This
expression is positive definite for the real (or hyperbolic
complex$^{11}$) version of NGT.  Here, we have taken into account
that $W_{[\mu,\nu]}$ is zero to lowest order of approximation
in the wave-zone.
\vskip 0.2 true in
\setsection\proclaim 7. {\bf Generation of Gravitational Waves in
the Linear Approximation} \par
\vskip 0.2 true in
The analysis of
the gravitational waves generated by the the GR source,
$T_{(\mu\nu)}$, in the linear approximation is well
understood$^{14}$.  Let us consider the possible existence of
radiation generated by the NGT sources, $S^\mu$ and
$T_{[\mu\nu]}$, in terms of their Fourier components $S^\mu({\bf x},t)$
and $T_{[\mu\nu]}({\bf x},t)$ in the linear approximation:
$$
S^\mu({\bf
x},t)=\int_0^\infty d\omega S^\mu({\bf x},\omega)
\hbox{exp}(i\omega t)+ CC, \sectionit $$ $$ T_{[\mu\nu]}({\bf
x},t)=\int_0^\infty d\omega K_{[\mu,\nu]}({\bf x},\omega)
\hbox{exp}(i\omega t)+ CC, \sectionit
$$
where $CC$ means complex conjugate.

In the wave-zone, we get
$$
\vert {\bf x}-{\bf x}^\prime\vert= r
- \hat {\bf x}\cdot {\bf x}^\prime +...
\sectionit
$$
We have $h_{[\mu\nu]}^{TT}=0$ and from (3.31), we obtain
$$
h_{[\mu\nu]}^{LT}=\partial_\nu\biggl(\int d^3x^\prime{S_\mu({\bf
x^\prime},\omega)\hbox{exp}[i\omega(t-\vert{\bf x}-{\bf
x}^\prime\vert)]\over \vert{\bf x}-{\bf x}^\prime\vert}\biggr)
$$
$$ -\partial_\mu\biggl(\int d^3x^\prime{S_\nu({\bf
x^\prime},\omega)\hbox{exp} [i\omega(t-\vert{\bf x}-{\bf
x}^\prime\vert)]\over \vert{\bf x}-{\bf x}^\prime\vert}\biggr).
\sectionit
$$

By using the expansion (7.3), we get
$$
h_{[\mu\nu]}^{LT}=e^{LT}_{[\mu\nu]}({\bf
x},\omega)\hbox{exp}(ik_\lambda x^\lambda)+CC,
\sectionit
$$
where
$$
e_{[\mu\nu]}^{LT}={1\over r}[k_\nu S_\mu({\bf k},\omega) -k_\mu
S_\nu({\bf k},\omega)]+O(1/r^2).
\sectionit
$$

When we substitute (7.6) into (6.4), we find that
$$
<t_{(\mu\nu)}>={k_\mu k_\nu\over
16\pi}[(e^{(\beta\gamma)}e^*_{(\beta\gamma)} -{1\over 2}\vert
e\vert^2)].
\sectionit
$$
Thus, the $e^{LT}_{[\mu\nu]}$ do not
contribute to the gravitational wave flux through the source
$S_{[\mu,\nu]}$, and this demonstrates that $h^{LT}_{[\mu\nu]}$
does not propagate.  This agrees with the conclusion, reached in
Sect.  4, that the spin $1^-$ sector does not propagate.  Since
the NGT pseudovector source $J^\mu=0$ to linear order, the gravitational
wave flux is the same to leading order as in GR.  This coincides with the
results obtained previously$^{1-3}$.  We observe that
$<t_{(\mu\nu)}>$ is {\it positive definite}, and the linear
approximation is physically consistent.
\vskip 0.2 true in
\setsection\proclaim 8. {\bf NGT on a Curved GR Background} \par
\vskip 0.2 true in
We shall expand the NGT field equations in
powers of $h_{[\mu\nu]}$, taking the symmetric part $g_{(\mu\nu)}$
to be an exact GR background, $g^{GR}_{\mu\nu}$.  To lowest order
in $h_{[\mu\nu]}$$^{19,1,2,4-7}$:
$$
g_{\mu\nu}=g^{GR}_{\mu\nu}+h_{[\mu\nu]}.
\sectionit
$$
The field equations with sources become in lowest order:
$$
R^{GR}_{\mu\nu}-{1\over 2}g^{GR}_{\mu\nu}R^{GR}=8\pi T_{(\mu\nu)},
\sectionit
$$
$$
D^\alpha F_{\mu\nu\alpha}+4{R_{GR}}^{\alpha \;\;
\beta}_{\;\; \mu \;\; \nu}h_{[\alpha\beta]}= -{4\over
3}W_{[\mu,\nu]}-{16\pi\over 3}S_{[\mu,\nu]}+16\pi T_{[\mu\nu]},
\sectionit
$$
$$
D^\nu h_{[\mu\nu]}=4\pi S_\mu,
\sectionit
$$
where $D^\alpha$ is the GR background covariant derivative and
$$
F_{\mu\nu\rho}=h_{[\mu\nu],\rho}+h_{[\nu\rho],\mu}+h_{[\rho\mu],\nu}.
\sectionit
$$
We observe that these equations are not invariant under the gauge
transformation (3.14), as is expected, for this invariance does
not occur in the theory.

Because of (3.17) and (8.5), we can rewrite Eq. (8.3) in the
form:
$$
D^\alpha D_\alpha h_{[\mu\nu]}+4{R_{GR}}^{\alpha \;\;
\beta}_{\;\; \mu \;\; \nu}h_{[\alpha\beta]}={\tilde
W}_{[\mu,\nu]}+ 16\pi\epsilon_{\mu\nu\alpha\beta}J^{\alpha,\beta},
\sectionit
$$
where
$$
{\tilde W}_{[\mu,\nu]}=-{4\over 3}W_{[\mu,\nu]}-{8\pi\over 3}
S_{[\mu,\nu]}+16\pi K_{[\mu,\nu]}.
\sectionit
$$
By taking the curl of (8.6), we obtain the field equation:
$$
D^\alpha D_\alpha F_{\mu\nu\rho}+D_{\{\rho}Q_{[\mu\nu]\}}=0,
\sectionit
$$
where $\{...\}$ denotes the curl. Moreover, we have
$$
Q_{[\mu\nu]}= 4{R_{GR}}^{\alpha \;\;
\beta}_{\;\;\mu \;\;\nu}h_{[\alpha\beta]}-16\pi\epsilon_{\mu\nu\alpha\beta}
J^{\alpha,\beta}.
\sectionit
$$
On the other hand, we could take the divergence of Eq. (8.6) and
get:
$$
D^\sigma D_\sigma {\tilde W}_\mu= -D^\nu(8{R_{GR}}^{\alpha\;\;
\beta}_{\;\; \mu\;\; \nu}h_{[\alpha\beta]}),
\sectionit
$$
where we have used the gauge condition:
$$
D^\mu {\tilde W}_\mu=0.
\sectionit
$$

Eq. (8.10) represents four {\it integrability conditions} that
must be satisfied, once the solutions for $g^{GR}_{\mu\nu}$ and
$h_{[\mu\nu]}$ are obtained.  The boundary conditions are imposed
on $g^{GR}_{\mu\nu}$ and $h_{[\mu\nu]}$ by demanding that
$g^{GR}_{\mu\nu}\rightarrow \eta_{\mu\nu}$ and
$h_{[\mu\nu]}\rightarrow 0$.

It has been proved by explicitly solving the system of six equations:
$$
D^\nu h_{[\mu\nu]}=4\pi S_\mu,
\sectionit
$$
$$
D^\alpha D_\alpha F_{\mu\nu\rho}=-D_{\{\rho}Q_{[\mu\nu]\}},
\sectionit
$$
that consistent solutions can be obtained for the six
$h_{[\mu\nu]}$, which satisfy the physical boundary condition of
asymptotic flatness at infinity$^{1-4}$.  It is important to
observe that any $h_{[\mu\nu]}$ solutions of (8.12) and (8.13)
must satisfy the primary equations (8.6). The same is true for any
${\tilde W}_\mu$ solution of (8.10)$^{4,20}$.

The skew contribution to the stress-energy pseudotensor,
$t^{(\mu\nu)}$, can be expressed in the form$^{6}$:
$$
t^{(\mu\nu)}_S={1\over 2}(F^{\mu\alpha\beta}F^\nu_{\alpha\beta}
-{1\over 12}g^{GR\mu\nu}F^2) +2\biggl({2\over
3}h^{[\mu\alpha]}g^{GR\,\nu\beta}W_{[\beta,\alpha]}
$$
$$
+{2\over 3}h^{[\nu\alpha]}g^{GR\,\mu\beta}W_{[\beta,\alpha]}
-{1\over
3}g^{GR\,\mu\nu}h^{[\alpha\beta]}W_{[\alpha,\beta]}\biggr)
-(3{R^{GR\,(\mu}}_{\gamma\alpha\beta}h^{[\nu)\alpha]}
h^{[\gamma\beta]}
$$
$$
-g^{GR\,\mu\nu}R^{GR}_{\gamma\alpha\delta\beta}h^{[\gamma\delta]}
h^{[\alpha\beta]})-D_\beta
D_\alpha(h^{[\alpha(\mu]}h^{[\nu)\beta]}).
\sectionit
$$
The GR Riemann tensor must have dimensions of $[\hbox{length}]^{-2}$ and
contains at most two time derivatives.  The dimensionless
quantity, $\dot M$, (where $M$ is the mass function) appears,
together with the quantity $\ddot M$, and therefore the Riemann
tensor behaves in the wave-zone as $\sim 1/r$.  The rate of energy
loss for an isolated body, in NGT, is given by
$$
{dE\over dt}=-R^2\int d\Omega t^{(0i)}\hat {n}_i,
\sectionit
$$
where the integration is over a sphere of radius $R$ in the wave-zone at
infinity and $\hat {n}_i$ is an outward pointing unit vector.  At
infinity, the first term in (8.14) is positive, while the second
term does not contribute, since in the flat spacetime limit
$W_{[\mu,\nu]}$ is zero to lowest order, as was shown
in Sect. 4. The third term goes to zero at least as $\sim 1/r^3$ and does not
contribute to the energy-momentum flux, while the fourth term has
the form:  $\partial_\alpha\partial_\beta
N^{[\mu\alpha][\nu\beta]}$, making no contribution.  Therefore,
the total energy-momentum flux in the wave zone is {\it finite and
positive definite}.

In the case of an axisymmetric, time-dependent solution of the NGT
field equations to first-order in $h_{[\mu\nu]}$ on a curved GR
background$^{2}$, and for the generalized Bondi, van der Burg,
Metzner and Sachs (BBMS) solution$^{3,21}$, the calculations show
that only the quadrupole GR radiation manifests itself in the
energy-momentum flux, for solutions that obey the asymptotic
flatness boundary condition at future and past null infinity.
This agrees with the result obtained in the linear
approximation in flat spacetime.  However, as in the case
of GR, it is not possible in the generalized BBMS solution to
interpret directly the nature of the near-zone NGT sources.  The
properties of these sources can only be inferred from multipole
moment expansions and their effects on wave solutions at future
and past null infinity.  It should be stressed that the NGT source
in the axisymmetric, reflexion symmetric solution represents a
generic source in NGT, for it possesses all the GR and NGT moments
of a general gravitational wave source.
\vskip 0.2 true in
\setsection\proclaim 9. {\bf Conclusions} \par
\vskip 0.2 true in
The proof of the physical consistency of NGT
does not depend on the existence of any form of gauge invariance
in the $g_{[\mu\nu]}$ sector, for there exist only two
conserved charges consistent with the two gauge invariances,
namely, diffeomorphism invariance and the $U(1)$ gauge invariance that
guarantees
the conservation of the charge $\ell^2$.  For an isolated
body, the physical boundary condition demands that in orthonormal, Cartesian
coordinates $g^{GR}_{\mu\nu} \rightarrow \eta_{\mu\nu}$ and
$g_{[\mu\nu]}\rightarrow 0$ as $r \rightarrow \infty$, which
in turn requires that for the static spherically symmetric NGT solution
in vacuum, $g_{[23]}=0$ to the lowest order of approximation.  In the higher
non-linear orders of approximation there is no direct contribution to
the flux of gravitational waves from the skew sector. It then
follows that NGT gives physically sensible answers for the radiation of
gravitational waves.

Axisymmetric time-dependent solutions have been obtained, which
show that the expected consistency of the field equations does
hold.  A solution to the axisymmetric case was derived to
first-order in $h_{[\mu\nu]}$, in an expansion about a curved GR
background$^{2}$.  The cross-coupling term between the background
GR Riemann curvature tensor and $h_{[\mu\nu]}$ does not cause any
inconsistency, provided the field equations are
correctly solved using the boundary conditions imposed on
$g_{\mu\nu}$, namely, that the solutions are asymptotically flat
at past and future null infinity$^{4}$.  The exact axisymmetric NGT
vacuum field equations were expanded in inverse powers of $r$, and
it was found that the gravitational wave flux of energy was
positive definite$^{3}$.  This result extended the proof of the
positivity of the gravitational wave flux in GR, obtained by Bondi
{\it et al.} and Sachs$^{21,22}$.

Because there are only two conserved
charges, $m$ and $\ell^2$, in NGT, it can be proved that there cannot
exist any direct contribution from $g_{[\mu\nu]}$, which has the form
of a retarded or advanced wave with the asymptotic behavior $g_{[\mu\nu]}
\approx 1/r$. This follows from the fact that $\ell^2$ has the dimensions
of a $[\hbox{length}]^2$, whereas $m$ (and the electric charge $e$ in
electromagnetism) has the dimensions of a length and can generate a
$1/r$ retarded wave solution and quadrupole radiation$^{23}$.

When the NGT charge $\ell^2=0$, the static spherically symmetric
solution only contains the mass $m$ and the theory becomes purely
geometrical. Test particles will follow geodesic equations determined
by the Christoffel symbols of NGT$^{24}$.

 From Eq. (8.2), we see that to
first-order in $h_{[\mu\nu]}$, a positive energy theorem follows
trivially using the methods of ref. 25.  However, further work
must be carried out to derive a general, rigorous positive energy
theorem in NGT.
\vskip 0.2 true in {\bf Acknowledgements} \vskip 0.2 true in This
work was supported by the Natural Sciences and Engineering
Research Council of Canada.  I am grateful to M. Clayton, N.
Cornish and P. Savaria for stimulating and helpful discussions.
\vskip 0.2 true in \centerline{\bf References} \vskip 0.2 true in
\item{1.}{N.  J. Cornish and J. W. Moffat, Phys.  Rev.  D{\bf 47},
4421 (1993).}
\item{2.}{N.  J. Cornish, J. W. Moffat, and D. C. Tatarski, Phys.
Lett.  A{\bf 173}, 109 (1993).}
\item{3.}{N.  J. Cornish, J. W.
Moffat, and D. C. Tatarski, University of Toronto preprint,
UTPT-92-17, gr-qc/9306033, 1992.
\item{4.}{N.  J. Cornish and J.
W. Moffat, University of Toronto preprint, UTPT-9401,
gr-qc/9401018
\item{5.}{T.  Damour, S. Deser, and J. McCarthy, Phys.  Rev.
D{\bf 45}, R3289 (1992).}
\item{6.}{T.  Damour, S. Deser, and J.
McCarthy, Phys.  Rev.  D{\bf 47}, 1541 (1993).}
\item{7.}{T.  Damour, S. Deser, and J. McCarthy, preprint
IHES/P/93/56,
gr-qc/9312030, 1993.}
\item{8.}{J. W. Moffat, Phys. Rev. {\bf 19}, 3554 (1979); J. Math.  Phys.
{\bf 21}, 1978 (1980); Found. Phys.  {\bf 14}, 1217 (1984).}
\item{9.}{J.  W. Moffat, Phys. Rev.  D {\bf 35}, 3733 (1987).}
\item{10.}{J.  W. Moffat and E. Woolgar, {\it ibid.} D {\bf 37},
918 (1988); Class.  Quantum Gravit.  {\bf 5}, 825 (1988).}
\item{11.}{For a review of NGT, see:  J. W. Moffat, Proceedings of
the Banff Summer Institute on Gravitation, Banff, Alberta, August
12-25, 1990, edited by R. B. Mann and P. Wesson (World Scientific,
Singapore, 1991), p. 523.}
\item{12.}{R.  B. Mann and J. W.
Moffat, J. Phys.  A {\bf 14}, 2367 (1981); {\it ibid}., {\bf 15},
1055(E) (1982).}
\item{13.}{A. Papapetrou, Proc. Roy. Irish  Acad. A {\bf 52}, 69 (1948);
M. Wyman, Can. J. Math. {\bf 2}, 427 (1950);
W. B. Bonnor, Proc. Roy. Soc. {\bf 209}, 353 (1951); {\bf 210}, 427 (1952);
J. R. Vanstone, Can. J. Math. {\bf 14}, 568 (1962).}
\item{14.}{S. Weinberg, {\it Gravitation and Cosmology} (New
York:  Wiley and Sons) p. 251, 1972.}
\item{15.}{R. B. Mann and J. W. Moffat, Phys.  Rev.  D
{\bf 26}, 1858 (1982).}
\item{16.}{R.  J. Rivers, Nuovo Cimento
{\bf 34}, 387 (1964); P. van Nieuwenhuizen, Nucl.  Phys.  B {\bf
60}, 478 (1973); E. Sezgin and P. van Nieuwenhuizen, Phys.  Rev.
D {\bf 21}, 3269 (1980).}
\item{17.}{This is the same result
obtained in:  M. Kalb and P. Ramond, Phys.  Rev.  D{\bf 9}, 2274
(1974) and V. I. Ogievetskii and I. V. Polubarinov, Sov.  J. Nucl.
Phys.  {\bf 4}, 156 (1967).  However, in the Kalb-Ramond formalism
the Lagrangian is explicitly invariant under (3.14); the linear
approximation to the NGT field equations should not be confused
with this formalism.}
\item{18.}{B.  Lautrup, K. Danske Vidensk.
Selsk.  Mat.  Fys.  Medd.  {\bf 35}, 1 (1967); N. Nakanishi,
Progr.  Theor.  Phys.  {\bf 35}, 1111 (1966); {\it ibid}., {\bf
38}, 881 (1967); N. Nakanishi and I. Ojima, {\it Covariant
Operator Formalism of Gauge Theories and Quantum Gravity}, World
Scientific, Singapore, 1990.}
\item{19.}{P.  F. Kelly, Class.  Quantum Grav.  {\bf 8}, 1217
(1991); {\bf 9}, 1423(E) (1992).}
\item{20.}{The no-go theorem used as a criticism of NGT, in ref.
7, does not apply, because there is no
direct skew field contribution to the radiation either at future
or past null infinity. Moreover, the theorem does not hold when
the constraints imposed by the primary equation (8.6) are taken
into account when solving Eq. (8.10) for ${\tilde W}_\mu$.
\item{21.}{H. Bondi, M. G. J.
van der Burg and A. W. K. Metzner, Proc.  Roy.  Soc.  A {\bf 269},
21 (1962); R. K. Sachs, Proc.  Roy.  Soc.  A {\bf 270}, 103
(1962).}
\item{22.}{P.  T. Chru\'sciel, M. A. H. MacCallum and D.
B. Singleton, preprint (1993).  In this work, it was shown that
the occurrence of terms like $r^{-j}\hbox{ln}^ir$ in the
Bondi-Sachs type method, due to the non-vanishing of the Weyl
tensor at Scri, does not lead to any serious difficulties in the
analysis of the geometry.  The Bondi-Sachs mass loss law continues
to hold, regardless of the occurrence of some high powers of
$\hbox{ln}r$ in the $1/r$ terms in the metric.}
\item{23.}{N. J. Cornish and J. W. Moffat, in preparation.}
\item{24.}{N. J. Cornish and J. W. Moffat, University of Toronto preprint,
UTPT-94-04, February 1994.}
\item{25.}{P.  Schoen and S. T. Yau, Phys.  Rev.  Lett.  {\bf 42},
547 (1979); E. Witten, Commun.  Math.  Phys.  {\bf 80}, 381 (1981);
J. M. Nester, Phys.  Lett.  A{\bf 83}, 241 (1981); R. Penrose, R. Sorkin
and E. Woolgar, Syracuse University preprint, 1993.}

\end